\documentclass[format=acmsmall]{acmart}\usepackage[]{graphicx}\usepackage[]{color}
\makeatletter
\def\maxwidth{ %
  \ifdim\Gin@nat@width>\linewidth
    \linewidth
  \else
    \Gin@nat@width
  \fi
}
\makeatother

\definecolor{fgcolor}{rgb}{0.345, 0.345, 0.345}

\usepackage{framed}
\makeatletter
 {\par\unskip\endMakeFramed%
 \at@end@of@kframe}
\makeatother

\definecolor{shadecolor}{rgb}{.97, .97, .97}
\definecolor{messagecolor}{rgb}{0, 0, 0}
\definecolor{warningcolor}{rgb}{1, 0, 1}
\definecolor{errorcolor}{rgb}{1, 0, 0}
\newenvironment{knitrout}{}{} 

\usepackage{alltt}
\renewcommand\footnotetextcopyrightpermission[1]{}

\usepackage[utf8]{inputenc} 
\usepackage{xfrac,xspace}
\usepackage{booktabs}
\usepackage{longtable}
\usepackage{multirow}
\usepackage{multicol}
\usepackage{amsthm}
\usepackage{caption}
\usepackage{subcaption}
\usepackage{xifthen}
\usepackage[]{graphicx}

\usepackage{adjustbox}
\usepackage{aeguill}

\newcommand{\enableTikz}[1]{}

\enableTikz{
\usepackage{tikz}
\usetikzlibrary{calc, chains, arrows, shadows, decorations.markings, decorations, fit, matrix, backgrounds, arrows.meta, positioning, patterns, shapes}
\usetikzlibrary{graphdrawing}
\usetikzlibrary{graphs}

\usetikzlibrary{external}
\tikzexternalize[prefix=./figures/,shell escape=-enable-write18]
\tikzset{external/system call={lualatex
\tikzexternalcheckshellescape -halt-on-error -interaction=batchmode
-jobname "\image" "\texsource"}}

\usegdlibrary{trees, layered}
\usegdlibrary{force}
\usegdlibrary{circular}

\newsavebox\gearbox

}

\newcommand{\powerset}[1]{\ensuremath{\mathcal{P}(#1)}\xspace}

\newcommand{\Verts}{\ensuremath{\mathcal{V}\xspace}}
\newcommand{\Edges}{\ensuremath{\mathcal{E}\xspace}}
\newcommand{\Labels}{\ensuremath{\mathcal{L}\xspace}}
\newcommand{\Graph}{\ensuremath{\mathcal{G}\xspace}}

\newcommand{\labelsOfVsymb}{\ensuremath{f_\Verts\xspace}}
\newcommand{\labelsOfEsymb}{\ensuremath{f_\Edges\xspace}}

\newcommand{\sizeOfV}{\ensuremath{\mathit{n}\xspace}}
\newcommand{\sizeOfE}{\ensuremath{\mathit{m}\xspace}}

\newcommand{\baseOfV}[1]{\ensuremath{\pi(#1)\xspace}}
\newcommand{\baseOfE}[2]{\ensuremath{\pi(#1,#2)\xspace}}

\newcommand{\GSL}{\texttt{GSL}\xspace}

\newcommand{\CASCAde}{European Research Council (ERC) Starting Grant CASCAde (GA n\textsuperscript{o} 716980)}

\setcopyright{none}
\acmDOI{00000/xxxx.xxxx}
\startPage{1}
\acmJournal{TOPS}
\acmVolume{99}
\acmNumber{0}
\acmArticle{0}
\acmMonth{0}

\IfFileExists{upquote.sty}{\usepackage{upquote}}{}
\begin{document}


\let\\\relax 

\title{\GSL: A Cryptographic Library for the strong RSA Graph Signature Scheme}

\author{Ioannis Sfyrakis}
\author{Thomas Gro{\ss}}
\affiliation{%
  \institution{Newcastle University}
  \department{School of Computing}
  \city{Newcastle Upon Tyne}
  \postcode{NE4 5TG}
  \country{UK}
}

%
%


%
%


\begin{abstract}
Current cloud and network infrastructures do not employ privacy-preserving methods to protect their assets. Anonymous credential schemes are a cryptographic building block that enables the certification of data structures and prove properties over their representations without disclosing the innards of their data structures in zero-knowledge.
The GRaph Signature (GRS) scheme devised by ~\cite{Gross:tz, Gross:2014fla} enables the certification and proof methods to sign infrastructure topologies represented as graph data structures and use zero-knowledge to prove properties over their certificates. As such, they represent a powerful privacy-preserving method that proves properties over a signed topology graph to another party without disclosing the blueprint of its topology.  In this paper, we report our efforts in designing, implementing and benchmarking a Graph Signature Library (\GSL). \GSL is a cryptographic library realized in Java that implements the graph signature scheme. \GSL enables application developers to create novel applications and services using graph signatures. We discuss the design decisions that were made during the development of the library. \GSL is based on a layered architecture that abstracts the implementation details of the underlying cryptographic scheme allowing the easy integration with other cryptographic schemes. We discuss a number of valuable lessons we learned during the development of the cryptographic library. We present the testing methodology that we followed for this library and how we ensure that there is not any information leakage of sensitive information. We evaluate the applicability of this library using graph topologies that represents a cloud infrastructure using a variety of configurations in terms of vertices and edges. The performance of the library is evaluated using benchmarks that show that the library has comparable performance to other RSA-based implementations.
\end{abstract}

\fancyfoot{}
\maketitle
\thispagestyle{empty}

\section{Introduction}
\label{sec:intro}

Anonymous credentials enable different parties to have privacy-protected communications in various domains such as cloud computing and IoTs. Design and implementation of anonymous credentials such as idemix~\cite{camenisch2002design} and U-Prove~\cite{paquin:uprove:2013} takes a lot of effort, mainly due to the complexity of the underlying protocols that are used.

The graph signature scheme devised by Gross~\cite{Gross:tz, Gross:2014fla} allows an auditor $A$ to certify infrastructure topologies such as a cloud infrastructure topology of a cloud provider (CP), and a prover $P$ that assures a verifier $V$ of policies regarding the infrastructure in zero-knowledge proofs without disclosing the blueprint of the infrastructure. For instance, a cloud tenant who acts as the verifier requests the cloud provider who acts as the prover to prove in zero-knowledge if her VMs are isolated or belong to a particular geographical location such as a country. The scheme is based on a signature scheme and corresponding honest-verifier zero-knowledge proofs of knowledge on graphs. Privacy-preserving techniques are used in a variety of domains from cryptocurrencies to cloud computing and Internet of Things (IoT).

Proving properties of an infrastructure that is represented as a graph without disclosing the details of the infrastructure in zero-knowledge has the potential of creating new computing paradigms where the privacy and confidentiality goals take central role.

In this paper, we propose \GSL, the \textit{Graph Signature Library}, a Java library for realizing the graph signature scheme. The library provides an API for creating different proof components and orchestrating the protocol stages for the graph signature scheme. The library provides an encoder that can encode arbitrary graph topologies using a graph encoding scheme according to the use case. For instance, we are using geo-location separation as an example of an encoding scheme. Each vertex label encodes a particular country that represents the location of the actual vertex.  Using a flexible and easy to understand API the developer is able to make choices that will not create an insecure implementation. 

Therefore, it is of paramount importance to ensure that the cryptographic library is developed in a way that follows the design goals we define.

\subsection{Contributions}
\label{subsec:contributions}
We offer the first realization of the graph signature scheme implemented in the Java programming language. We present the architecture of the library that uses a novel approach on creating orchestrators to coordinate reusable components for the fulfillment of zero-knowledge proofs. We showcase how design patterns can be leveraged to design anonymous credentials libraries and the importance of creating a specification document for cryptographic algorithms that matches the implementation of low-level cryptographic primitives. We discuss the variety of testing methods of the library in terms of information flow, static analysis and coverage. We develop a number of benchmarks for each stage of the cryptographic scheme to evaluate the performance of the library in terms of different encryption keys and size of the graph topology. The performance evaluation measures the execution time of the proof components during the issuing, proving and verification stages. In addition, we discuss lessons learned throughout the development of the first cryptographic library that implements the GRS scheme, which application developers and users can benefit.

\textbf{Outline:}
We organize the rest of the paper as follows. We describe the graph signature scheme in section~\ref{sec:graphsignatures}. Section~\ref{sec:design_principles} discusses the design principles driving the architecture of the GSL.
In section~\ref{sec:design}, we describe the design of the GSL and how we devised its architecture. Next, we present the details of the implementation of the GRS library in section~\ref{sec:implementation}. In section~\ref{sec:evaluation}, we present the results from the performance evaluation and the testing of the library. Section~\ref{sec:relatedwork} discusses related work to this paper. Section~\ref{sec:conclusion} states conclusions and outlines future research threads.

\section{Graph Signatures}
\label{sec:graphsignatures}

Graph signatures is a novel signature scheme where a graph structure can use zero-knowledge proofs to prove properties such as isolation to other parties without revealing any other information apart from the truth of the property. We now discuss required preliminary material for graph signatures.

\subsection{Preliminaries}

\textbf{Assumptions.}
The graph signature scheme is based on the Strong RSA assumption~\cite{rivest1978method,fujisaki1997statistical}, which operates on the special RSA group. The set $QR_N$ is the cyclic subgroup of Quadratic Residues of a special RSA group with modulus $N$.

\textbf{Integer Commitments.}
When the Pedersen commitment scheme~\cite{Pedersen1991} operates in a special RSA group and the committer does not know the factorization of modulus, then the commitment scheme can be used to commit to integer of arbitrary size~\cite{damgard2001}.

\textbf{Known Discrete-Logarithm-Based, Zero-Knowledge Proofs.}
There are several known results on proving statements about discrete logarithms. For instance the proof of knowledge of a discrete logarithms modulo~\cite{Schnorr1991} a prime or a composite~\cite{damgard2001, fujisaki1997statistical}.

\textbf{Camenisch-Lysyanskaya signatures.}
The graph signature scheme uses the Camenisch-Lysyanskaya (CL) scheme in a strong RSA setting~\cite{camlysy2003}. It is used to sign hidden messages and prove knowledge of a signature.

\textbf{Camenisch-Gro{\ss} encoding.}
The Camenisch-Gro{\ss}~\cite{camenisch2012efficient} encoding manages to encode multiple binary and finite set values into a single message on the CL message space. The main idea is to represent binary and finite set attributes using prime numbers. The scheme uses divisibility and co-primality to establish whether an encoded attribute value exists or not in a credential. The credential issuer certifies the association between the attribute value and the prime number.

\textbf{Graph Encoding.}
Table~\ref{tab:not-graphenc} provides and overview of the graph encoding notation. Graphs with finite set of vertices and undirected edges or directed arcs are encoded alongside a finite set of vertex and edge labels~\cite{gross2015signatures}. Each vertex in the graph is represented by a prime number in credentials and proofs. All prime numbers in this set are pair-wise different. Using the same approach each label in the graph is assigned a prime number, which acts as a prime representative. All prime representatives in this set are also pair-wise different.

Integer commitments and CL-signatures encode vertices and edges into their exponents, which makes the exponents accessible to proofs over the exponents. The association of the bases with the vertices and edges is randomized. The encoding of a graph involves the following process. The product of the prime number associated with the vertex or edge and the prime representatives of the associated labels is calculated and then encoded as an exponent of a vertex or edge base.

\begin{table}[t]
\centering
\caption{Notation for graph encoding}
\label{tab:not-graphenc}
\begin{tabular}{@{}ll@{}}
\toprule
$\Verts$ & Finite set of vertices \\
$\Edges \subseteq (\Verts \times \Verts)$ & Finite set of edges \\
$\Graph=(\Verts, \Edges, t_\Verts, t_\Edges)$ & Graph  \\
$\Labels_\Verts, \Labels_\Edges$ & Finite sets of vertex and edge labels \\
$\labelsOfVsymb : \Verts \rightarrow \powerset{\Labels_\Verts}$ & Labels of a given vertex  \\
$\labelsOfEsymb : \Edges \rightarrow \powerset{\Labels_\Edges}$ & Labels of a given edge  \\
$\sizeOfV = |\Verts|, \sizeOfE = |\Edges|$ &  Number of vertices and edges \\ \bottomrule
\end{tabular}%
\end{table}

\textbf{Signatures on Committed Graphs.}

The graph signature scheme creates commitments over a graph and proves predicates over the graph structure with a zero-knowledge proof of knowledge.

\section{Design Principles}
\label{sec:design_principles}
In this section we outline the main design principles that guide us throughout the design and development of the graph signature library.
Even though, design and implementation of a cryptographic scheme is considered a difficult endeavor having a specific set of design principles to guide us throughout the development of the library is always helpful and produces better quality software. 

Before we begin discussing the design principles for \GSL we motivate our reasoning behind choosing the Java programming language for developing the library. First, the Java programming language caters for development of large and complex software projects by using a large standard library, garbage collection and memory management without using pointer logic. Second, Java can be run on a plethora of operating systems and hardware platforms. Third, there is a number of cryptographic libraries already implemented in Java. For instance, idemix is a library that implements anonymous credentials in Java. In addition, Java provides the Java Cryptography Architecture (JCA)~\cite{jca} API with the Java Development Kit (JDK) where multiple providers can be used such as the Bouncy Castle cryptographic library~\cite{bouncycastle:api:2019}. Finally, Java provides the Java Native Interface (JNI)~\cite{jni} which enables developers to call native code that is optimized to the specific hardware and software. This approach can further optimize the performance of cryptographic operations. In addition, using JNI  other libraries and applications that use native code can be integrated with the Java code. 

\textbf{Security - hard to misuse API} One salient feature that cryptographic libraries need to address is giving the user of the library minimal ways of using the API without misusing the parameters for each function call. This enables the user to avoid using the API only with the intended way. For instance, there are not many options in the library both in the operation of the graph signatures and in the encoding of graphs.

\textbf{Usable API}
The library provides a simple API that developers can use. We calculate a number of complexity metrics to quantify the usability of the cryptographic library.

\textbf{Extensible}
\GSL makes it accessible to developers to add new cryptographic schemes and group implementations into the lower level. There is also the ability to add new graph encoding schemes according to the context the library operates. 

\textbf{Parameterizable}
Allows for easy customization of parameters that govern the operation of the cryptographic library. This means that changing the system parameters both for the key generation, graph encoding and signature scheme is straightforward. We use configuration files that exist outside of the main source code to further improve the manageability of the parameters.

\textbf{Agile}
Ability to deprecate older schemes or algorithms that are broken or inefficient. In addition, a developer can use the \GSL API to create graph signatures using different settings such as different key length and topology graph configuration. 

\textbf{Interoperable}
The library is available to run in many platforms and can exchange messages with different platforms. Using Java programming language ensures that the library can be used in a variety of platforms. 

\textbf{Modular} The library encompasses small components that use design patterns and realize proof components and orchestrators for the issuing, proving and verifying stages of the graph signature scheme.

\textbf{Abstraction} \GSL incorporates abstractions over cryptographic groups for $QR_N$, the special RSA group with modulus $N$ and  abstractions over generators and group elements anticipating that in the future we can add more group and group element abstractions such as configurations for elliptic curves and bilinear pairings.

\textbf{Safe defaults} The library embeds safe defaults with 2048 RSA key lengths and appropriate default parameters for the graph encoding scheme and the signature scheme. Using safe defaults developers can improve their development time integrating \GSL with their own application and services. 

\textbf{Well documented and specified}
The library is well documented by providing detailed Javadocs that explain the behaviour of classes and methods. In addition, a detailed specification document~\cite{gross2018specification} is provided that details the low level cryptographic algorithms needed to realize the graph signature scheme.

\textbf{Performance} A detailed performance evaluation is executed to show if the library performs in par with other RSA-based cryptographic schemes.


\textbf{Reusable} We expect that components developed can be used in future implementations and support different cryptographic primitives such as elliptic curves or reusing proof components for the zero-knowledge proof of knowledge.

\section{Design and Architecture}
\label{sec:design}

In the following we discuss the design of the cryptographic library and motivate the design choices for this library. First, we outline the main layers of the architecture. Second, we discuss the graph encoding process and the design patterns we used to realize an abstract way of encoding graphs that employ an arbitrary encoding scheme. 

\subsection{\GSL Design}
We outline the design for the \GSL in four conceptual layers from bottom to top. The first layer is comprised of low-level primitives such as functions that  generate random numbers and primes, compute the Chinese remainder theorem and the Jacobi symbol etc. The second layer consists of functions related to supporting the graph encoding scheme and configuring the \GSL. The third layer includes the functionality that realizes the graph signature protocol employing the required parties such as the signer, recipient, prover and verifier. Finally, the fourth layer consists of the orchestration and communication layer that coordinates the different parties and sends messages between the parties during the execution of a proof instance for the graph signature scheme. 
 
\subsubsection{Layer 1 - Low-level primitives}
The first layer of the \GSL consists of the main cryptographic primitives used for the realization of the graph signature scheme. We opted to implement all the required cryptographic primitives from scratch without integrating third-party libraries and code. This results in self contained components that implement the required low-level cryptographic operations for the graph signature scheme. 

There are seven main computations that are addressed in this layer. First, the special RSA modulus is generated for the graph signature schemes. Second, the random safe prime are generated which are required for the special RSA modulus. Third, the commitment group and its generators are computed. Fourth, the computations related to group structures and their group elements when we know factorization of the modulus or not. Fifth, another function that belongs to this first layer is the Chinese Remainder Theorem (CRT) computation. Sixth, the Extended Euclidean Algorithm (EEA) is computed in this layer. Finally, the Jacobi symbol is computed in this layer, which evaluates if a group element is part of $QR_N$.     

\subsubsection{Layer 2 - Encoding and Configuration}
The second layer deals with the graph encoding, configuration, naming and storage of the \GSL. The encoding functionality in this layer focuses on how the encoding scheme is represented and applied to a particular graph. Each encoding scheme is configured via an external file which specifies the structure of the encoding scheme. The library is also configured using a main configuration file that specifies all the required parameters for the graph signature scheme. The parameters required are presented in a table in the specification document~\cite{gross2018specification}. Alongside this layer is the functionality that provides the naming for the different elements of the graph signature scheme. The naming facilities are used in the storage functionality for reading and writing members of a proof that a party during the execution of the graph signature scheme creates or consumes. 

\subsubsection{Layer 3 - Graph Signature Protocol}
The third layer realizes the main functionality for the graph signature scheme and especially the related parties. There are four parties for the graph signature scheme. First, the signer is the party to compute the key generation and the signing of committed graphs. Second, the recipient is responsible for initiating the signing process in the interactive protocol and completing the signature for the graph. Third, the prover main tasks are to compute the proof of possession and vertex and edge decomposition of the graphs alongside the prover's part of the pair-wise different proofs. Finally, the verifier aims to verify zero-knowledge proof of knowledge originating from the prover with respect to the signer public key and the policy predicate for what statement to verify. 

In addition to the parties in the graph signature schemes there are proof components that compute parts of the proofs. In an interactive setting there is proof component for each member of the proof instance. For instance, one component computes the prover side of the proof that pertains to the component and the other component computes the verifier side of the proof. There are four types of the proof components. First, are the proof components that evaluate if the key generation elements were computed correctly. Second, the proof components related to proving and verifying commitments for the issuing and proving stages. Third, the pair-wise difference proof components for proving pair-wise difference over vertices. Finally, the proof components related to proving possession of a graph signature.

\subsubsection{Layer 4 - Orchestration and Communication Layer}
This layer focuses on the orchestration of zero-knowledge proofs of knowledge and the communication of messages to between members of the proof protocol. 
The orchestrator components act as coordinators that manage the related proof components for a particular proof protocol. There are orchestrators that manage the execution of the issuing the graph signature and proving properties over the graph. Each orchestrator coordinates the execution of its side and also the message to send to the other party of the zero-knowledge protocol. For instance, during the proving and verifying stages of the graph signature scheme there is an orchestrator that coordinates the prover components and another orchestrator for the verifier components. 

In addition to the orchestrators, there are components that focus on the communication between the parties during the execution of protocol stage. Depending on how the communication needs to be established, a messaging component can be wired to send the protocol message using different methods, which can be synchronous or asynchronous. So one party will send its message while the other one will be waiting for the message and will send back the reply after processing the message and executing the required computations.

\subsubsection{Design Patterns}
We used a number of design patterns~\cite{pree1995design} to further improve the modularity and flexibility of the library. Using design patterns for the development of cryptographic libraries creates reusable software which act as building blocks for more complex system architectures~\cite{gamma1993design}. Especially when the cryptographic library needs to be integrated with other systems or libraries. By using design patterns we were able to communicate the structure and behaviour of the software implementation in an abstract way. The stakeholders of the project, which may not have a large experience in software engineering, improved their understanding of the intended software design. This has also been witnessed in large-scale software development projects~\cite{schmidt1995design} in industry. The main design patterns for the design process of the \GSL include the Proxy, the Facade, the Abstract Factory, Iterator and the Dependency Injection (DI) design pattern~\cite{prasanna2009dependency}.

The Proxy pattern was used to be able to switch to alternative communication protocols for the execution of the two party interactions for the participants of the graph signature scheme first for issuing a graph signature and then to prove properties to a remote verifier. Currently, the communication channel between the two parties can be either socket-based or HTTP-based. There is also the option of a local channel mainly to facilitate the testing of the library. The local channel mode can only be used for the testing of the library. The main advantage of the Proxy pattern is the hiding of the implementation details from the client, which makes the future change, test and reuse of the implementation easier.

The Facade pattern provides a unified interface for the low level cryptographic algorithms implemented in the library. This patterns supports loose coupling and hides the complex details by exposing a simple interface to client code. Using this approach different subsystems can be used anticipating future change in the library. For instance, we intend to add an Elliptic Curve subsystem for the graph signature library.

Another design pattern used in the \GSL is the Factory pattern which delegates the decision of which class to instantiate encapsulating the selection logic. This pattern enables the application developer that uses this library to choose which class will be used for the application. For instance, in \GSL we can select which component will implement the required low-level number theoretic algorithms. Currently, we instantiate the RSA-based component when we need to perform low-level computations. In the future, we anticipate another component that will implement the required Elliptic Curve or Bilinear Pairing algorithms, which the application developer can select if it is needed.

\begin{figure}[tb]
	\centering
	\includegraphics[width=.45\textwidth]{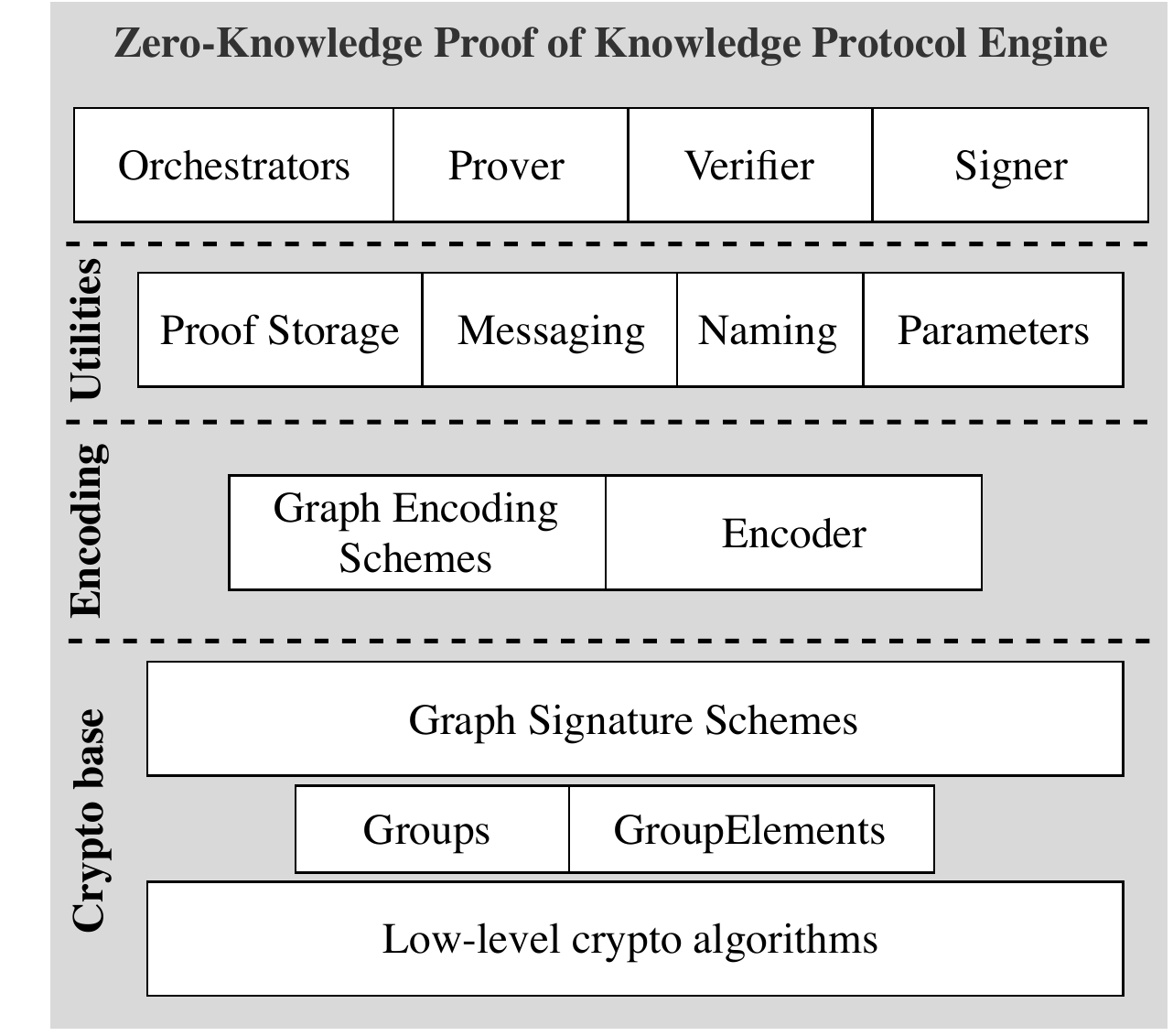}
	\caption{Overview of the \GSL architecture}
	\label{fig:grslib-architecture}
\end{figure} 

Iterators is a design pattern that provides a mechanism to access elements of a composed object in a sequential way without exposing its underlying representations. This pattern was used to traverse the base collection when we computing zero-knowledge proof protocols during the issuing, proving and verifying stages of the graph signature scheme. The added facility of the iterators enable different traversal schemes according to the computation context. For instance, when we are computing the pre-signature Group Element $Q$ we traverse both vertex and edge bases creating only one iterator instance. When we are computing witness randomness for the proof of possession we require only to traverse the vertex bases or the edge bases since the witness randomness will have different canonical name for vertices and edges. This way we conveniently store and retrieve the message randomness as required by the proof protocols.

Dependency Injection is used in the \GSL when instantiating a new orchestrator which coordinates the communication channel and the interactions between the proof components required for realizing a particular proof protocol. By using constructor dependency injection in this instance, we promote loose coupling between components and it can be easier to extend the orchestrator with new types of communication channels in the future if needed.

While designing the different components of the library we realized that there should be an entity that would coordinate the components that realize a zero-knowledge proof protocol and the communication between the two parties in each stage of the protocol. For instance, during the issuing stage we would need to coordinate the protocol and the interactions between the issuer and the recipient and during the proving stage between the prover and verifier. We created a design structure that first coordinates the execution of the proof components according to the particular proof protocol and at the same time managing the communication of the messages to the other party such as sending proof signature and nonces. We call this design entity an orchestrator.

\begin{figure}[tb]
  \begin{subfigure}[b]{0.4\textwidth}
    \includegraphics[width=\textwidth]{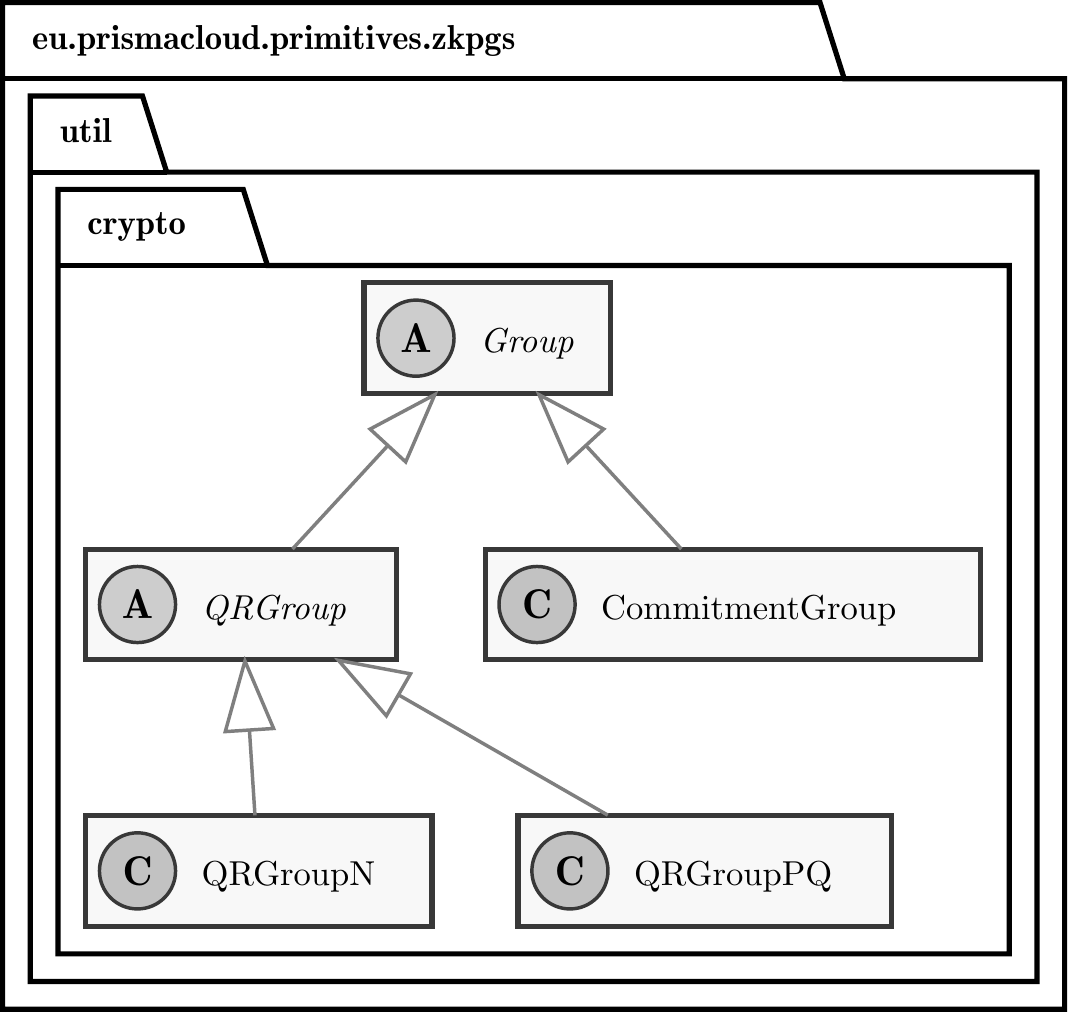}
	\caption{The class hierarchy of Groups in the \GSL}
	\label{fig:groups}
  \end{subfigure}
  \hfill
  \begin{subfigure}[b]{0.5\textwidth}
    \includegraphics[width=\textwidth]{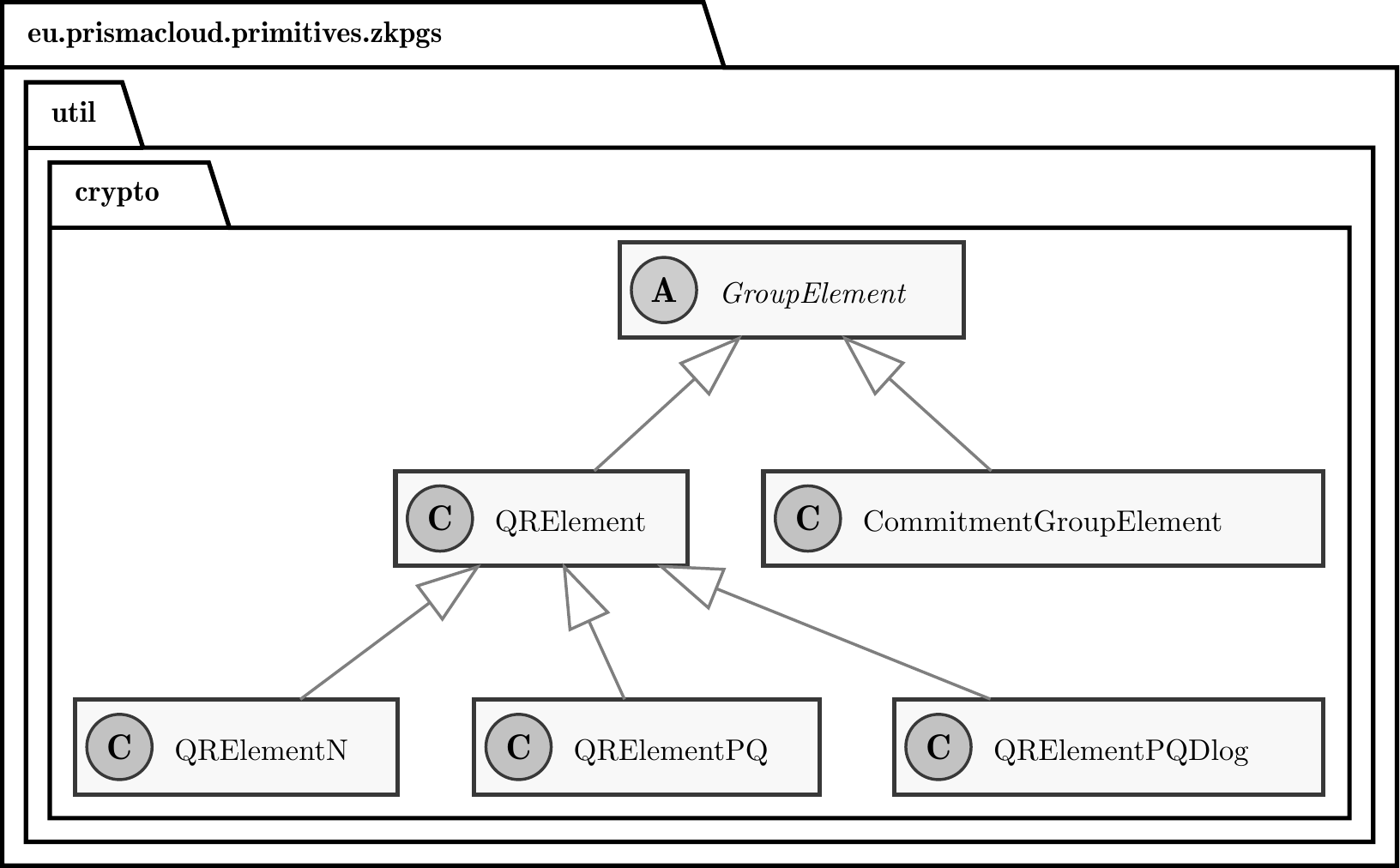}
	\caption{The class hierarchy of GroupElements in the \GSL}
	\label{fig:groupelements}
  \end{subfigure}
  \caption{The hierarchy of Groups and GroupElements in the \texttt{crypto} package.}
\end{figure}

\subsection{\GSL Architecture}
Figure \ref{fig:grslib-architecture} depicts the building blocks of the graph signature library. The lower level components comprise the cryptographic basis upon the graph signature scheme is realized. The components on the above layer focuses on the encoding of arbitrary graphs topologies. The utilities layer introduces auxiliary components that are used throughout the library and provide storage, naming, parametrization and communication functionality. The higher level components are part of the zero-knowledge protocol engine that aims to coordinate the interactions between the different parties and the execution of the proof protocols.

\paragraph{\textbf{Crypto base}} The lower layer introduces the main cryptographic algorithms required to realize the graph signature scheme. These are shown in Figure \ref{fig:low-level-crypto-algos}. This layer includes the groups and group elements for the graph signature scheme alongside the key generation algorithm.  Another component in this layer is related to the graph encoding schemes that the library requires to encode a graph. The main low-level cryptographic algorithms for the \GSL include the Chinese Remainder Theorem (CRT) algorithm, the Jacobi symbol, the Extended Euclidean  algorithm (EEA), the special RSA modulus and the algorithm for generating safe primes. The specification for each of these low-level cryptographic algorithms can be found in the following technical report~\cite{gross2018specification}.


Different groups are supported for the graph signature scheme depending on the user knowing the factorization of the modulus and using this information to speedup computations. We design the group components in a hierarchy which can support other groups needed for future extensions to the graph signature scheme. Using this approach we create an elegant design to support multiple group and group element specifications using a hierarchy, which can be reused and extended. For instance, when we don't know the factorization of the modulus then all computations are executed in the Quadratic Residues group $Z^*_n$ (\texttt{QRGroupN}). When we know the modulus factorization we work on the \texttt{QRGroupPQ}. The \texttt{CommitmentGroup} is used to constrain the commitment computations in a particular group.

Regarding group elements supported for the graph signature scheme, they include the \texttt{QRElementN} for using elements that are part of the \texttt{QRGroupN} when we don't know the factorization of the modulus. When we know the factorization of the modulus we use the \texttt{QRElementPQ}. The \texttt{QRElementPQDlog} element is used when we know the discrete logarithms and the modulus factorization. Using this approach we can speedup the computations. Commitments use the \texttt{CommitmentGroupElement} for these computations.

%

\begin{figure}[tb]
	\centering
		\includegraphics[width=15cm]{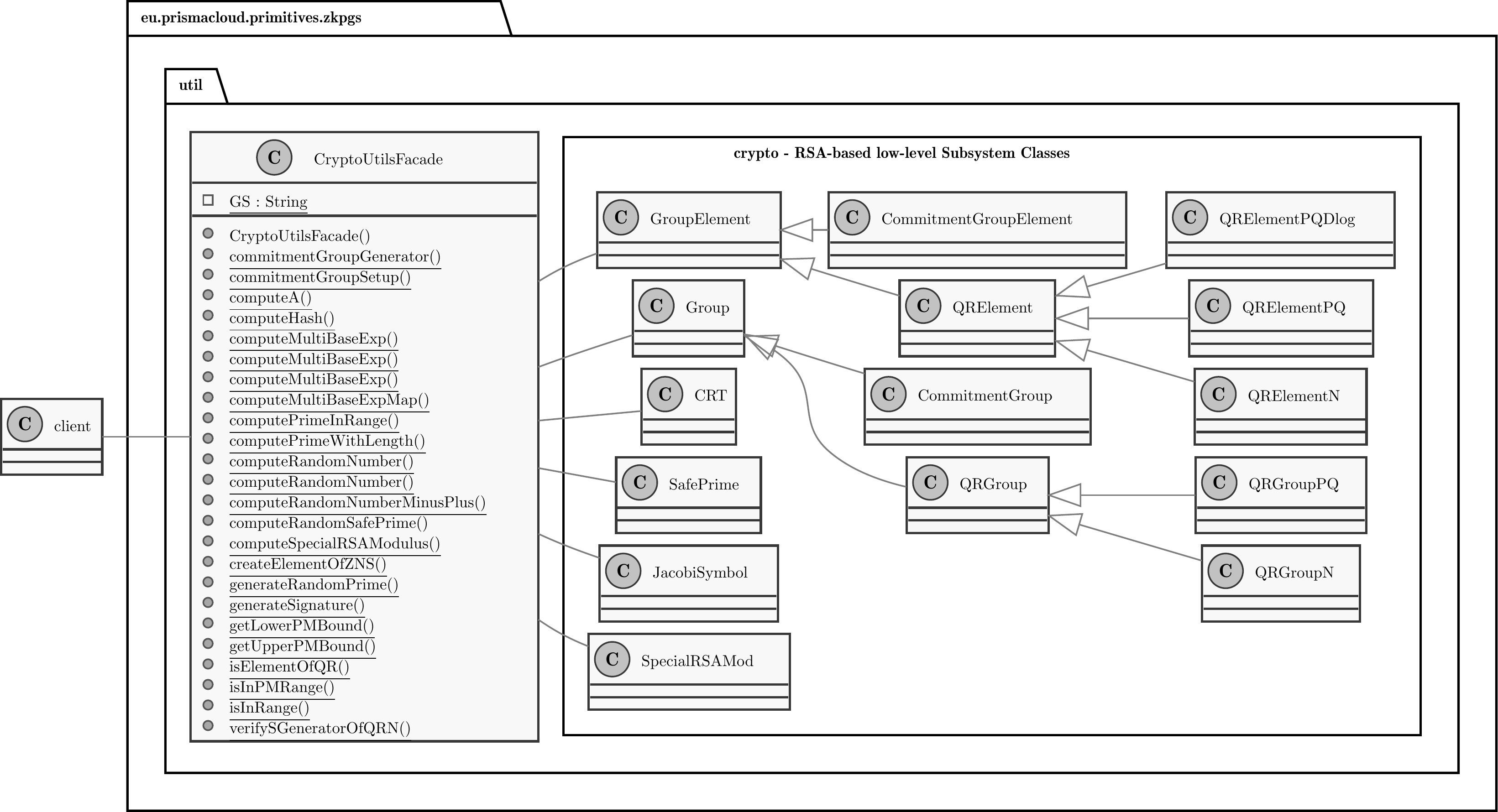}
	\caption{Using the facade design pattern for the RSA-based subsystem for the low-level cryptographic algorithms and groups}
	\label{fig:groupelements}
\end{figure}

Currently, the graph signature scheme works only on RSA-based creation of graph signatures. We consider the integration of graph signature scheme with elliptic curves as future work. The way we have designed the graph signature library allows us to incorporate elliptic curves as a subsystem in the crypto base that would focus on cryptographic algorithms for pairings and elliptic curves. This research thread will give graph signatures the ability to decrease the time of generating keys, since elliptic curves require a smaller key length to support the same security level as RSA-based schemes. Graph signatures based on elliptic curves and pairings would be more efficient and could be used in a variety of scenarios where efficiency is required such as embedded devices and the Internet of Things (IoT).

\begin{figure}[t]
	\centering
		\includegraphics[width=9cm]{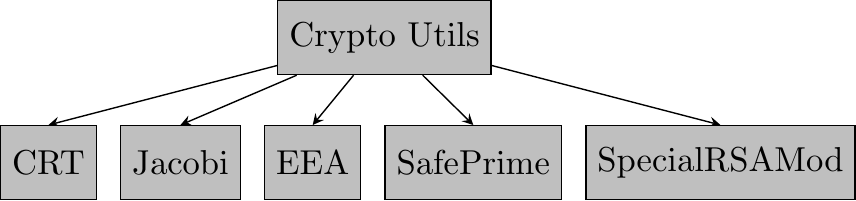}
	\caption{The low-level cryptographic algorithms implemented for the realization of the graph signature scheme}
	\label{fig:low-level-crypto-algos}
\end{figure}

\paragraph{\textbf{Utilities}}
The utilities layer includes components that assist the library to realise the zero-knowledge proof of knowledge proofs for the graph signature scheme. The main components in this layer are the \texttt{ProofStorage}, the \texttt{Naming} and \texttt{Messaging}. The goal of the first component is to provide a storage facility for each member of the proof protocol that can store and retrieve data related to the proofs. The second component introduces a naming scheme based on Uniform Resource Names (URNs)~\cite{daigle:urn:2002} that each party can use to uniquely identify the information stored in the \texttt{ProofStorage} and retrieve it for subsequent proofs. The third component provides the communication interface that each party can use during the protocol execution to send the required proof data to the correspondent party.

\paragraph{\textbf{ZKPoK Protocol Engine}}
This layer includes the components required for realizing the zero-knowledge protocols and proofs for the graph signature scheme. There are four components that resemble the different parties that need to interact with each other and fulfill the requested proof. The \texttt{Signer} component is responsible for the key setup and the signing of the committed graphs. The \texttt{Recipient} component initiates the signing process and completes the graph signature. The \texttt{Prover} component is responsible for the proof of possession for the graph signature. The \texttt{Verifier} component aims to verify zero-knowledge proofs of knowledge originating from the \texttt{Prover} component taking into account the signer public key and the proof request. The \texttt{Orchestrators} of each party coordinate the execution workflow of the proof protocol for the graph signature scheme.

\subsection{Encoding A Graph}

\begin{figure}[t]
	\centering
		\includegraphics[width=13cm]{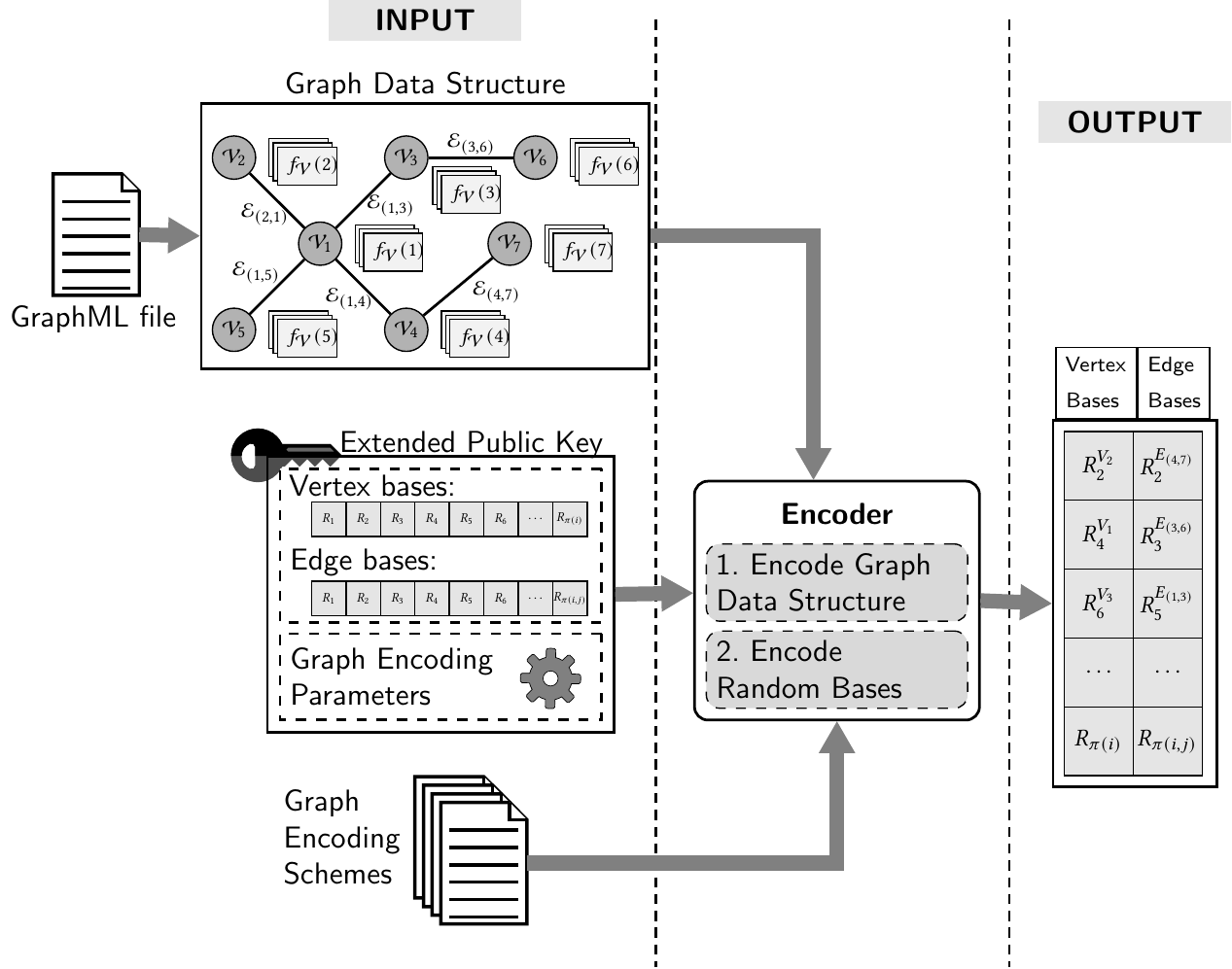}
	\caption{The graph encoding workflow}
	\label{fig:graph-encoding-workflow}
\end{figure}

In this subsection, we outline the process of encoding a graph and the design choices made during the development of the library. The graph encoding is based on the prime encoding introduced by Gro{\ss} and Camenisch~\cite{camenisch2012efficient}. The workflow of encoding a graph in \GSL is depicted in Figure~\ref{fig:graph-encoding-workflow}. The graph encoding workflow requires three inputs. The initial input is the GraphML~\cite{brandes2001graphml} file, which specifies the vertices, edges and labels of the topology graph. The file is imported by the library and instantiates an interim graph data structure to represent the GraphML file. The graph data structure is the first input to the encoder component. The second input is the extended public key that was previously generated during the extended key generation process. The extended public key consists of the bases for vertices and edges according to the graph encoding parameters. The third input is the graph encoding scheme that contains the information on how to encode the graph. For instance, if we are encoding graphs with geolocation information, then the encoding scheme contains the mapping of labels for a particular country to a corresponding prime number as discussed in~\cite{Gross:2014fla}.

\begin{figure}[t]
	\centering
		\includegraphics[width=15cm]{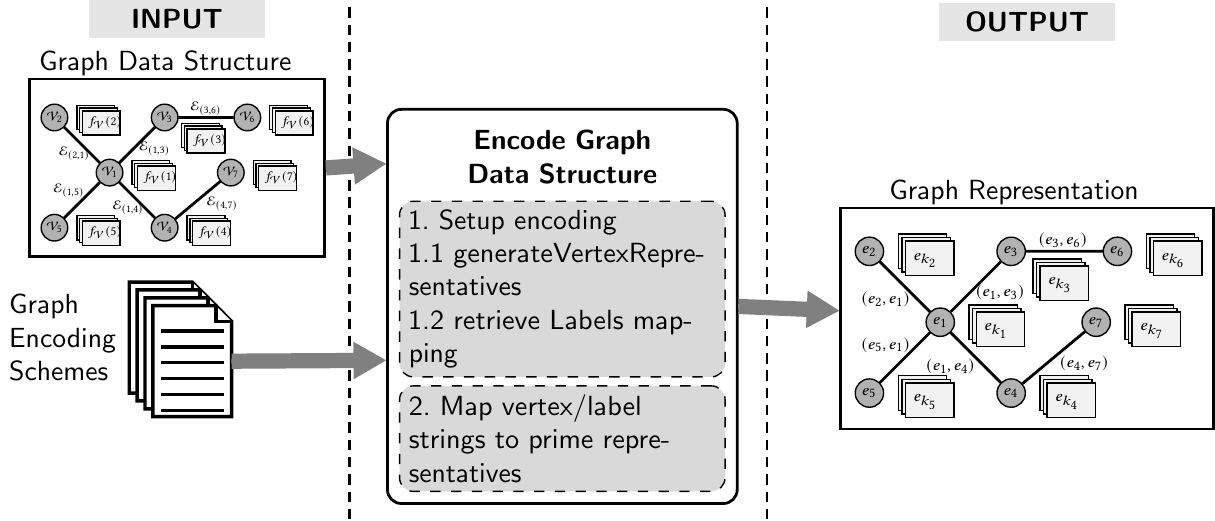}

	\caption{Stage 1: The process of encoding the graph data structure}
	\label{fig:graph-data-structure-encoding-process}
\end{figure}

The encoder component is responsible for encoding the created graph data structure and for encoding random bases from the extended public key. The process of encoding the graph data structure is illustrated in Figure~\ref{fig:graph-data-structure-encoding-process}. The steps for encoding the graph data structure involves two main steps. During the first step the encoding is setup and we generate vertex representatives and we retrieve the mapping of the labels from the graph encoding scheme. The second step we map vertex and label strings to prime representatives which will be passed to the second phase of the encoding. The output of this phase is the graph representation. This data structure references a particular vertex and label to particular prime representative. The graph representation created is passed to the second phase of the encoder.

\begin{figure}[t]
	\begin{adjustbox}{center}
	\centering
		\includegraphics[width=14cm]{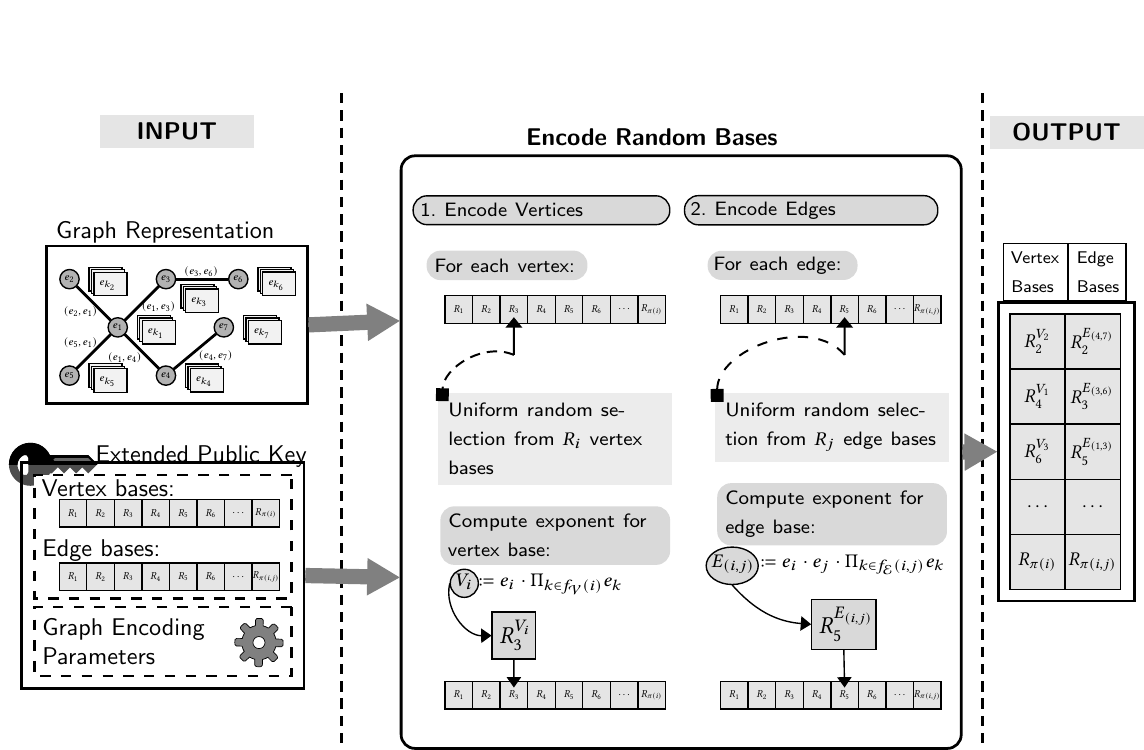}
	\end{adjustbox}
	\caption{Stage 2: The process of encoding random bases}
	\label{fig:random-bases-encoding-process}
\end{figure}

During the second phase of encoding a graph we input the graph representation and the extended public key. The process of this phase is depicted in Figure~\ref{fig:random-bases-encoding-process} The goal of this phase is to encode vertices and edges on uniformly random selected bases from the extended public key. We first encode vertices and we encode each vertex by first uniformly random selecting a vertex base $R_{\baseOfV{i}}$ and then computing the exponent for the selected vertex base as depicted in Figure~\ref{fig:random-bases-encoding-process}. A similar process is followed for encoding edges. For each edge we select a random edge base $R_{\baseOfE{i}{j}}$ and then we compute the exponent for the edge base. The final output of the encoding process is the encoded random bases for vertices and edges that hold the exponent for a random vertex or edge.


\subsection{Lessons learned}
During the iterative process of designing the architecture of \GSL we learned a number of valuable lessons, which are of further interesest to the developers of cryptographic libraries and to the research community.
We first discuss what we learned from creating a conceptual layered design and using an iterative pattern-based design process for the \GSL.

Designing complex software systems using a hierarchy of different levels of abstraction which incorporate encapsulation and information hiding results in quality architecture designs~\cite{neumann:hierarchy:1986,foss:multi:2004}. Thus, during the design process of the cryptographic library we separated the architecture of the library in different layers according to the goal of each set of components. 

We opted for using design patterns to ensure that we use proven solutions for problems that arise during the design and implementation of cryptographic libraries. Our main aim was not to reinvent the wheel.  Developing a new cryptographic library from scratch is a challenging endeavor and leveraging design patterns can ameliorate the difficulty of the task. Even though not every problem can be solved with a design pattern it was helpful to know that there are options that we could amend to our design and that we can also introduce our own specific design elements to the architecture of the library.

\section{Implementation}
\label{sec:implementation}
In this section, we discuss the implementation of the graph signature library and provide further details on components of the architecture.
First, we discuss how the key generation functionality was implemented. Second, we outline the classes that are part of the utility package and provide auxiliary methods to the \GSL. Third, we describe how the graph encoding was implemented. Finally, we discuss the classes and interfaces that are part of the core protocol layer.  

Before we start discussing the details of the implementation we briefly motivate the reasons behind choosing the version of Java. For the implementation of the graph signature scheme we used version 1.8 of Java to ensure compatibility with other applications and libraries that would like to make use of the graph signature library. The implementation of BigIntegers in Java provide a solid base upon to implement cryptographic 

\paragraph{Configuration} The graph signature library can be parameterized in two ways. The first is using a JSON\footnote{JavaScript Object Notation, http://www.json.org} file that holds the configuration of the parameters to realize the graph signature scheme. The list of parameters include system and group parameters required for the key generation and proofs for issuing and verifying. In addition, a separate file includes the parameters for the graph encoding scheme. All these parameters are defined in the library specification document.

\paragraph{Library specification} We implemented the library in close tandem with the library specification document~\cite{gross2018specification}. This way the implemented cryptographic algorithms and proof protocols closely match those defined in the specification document. In addition, the specification document can provide the basis for implementing the graph signature scheme in a different language such as C++ or Python.

\subsection{Key Generation}
The key generation functionality in the graph signature library is implemented using a range of classes and interfaces. The main classes that implement the key generation is the \texttt{SignerKeyPair} and the \texttt{ExtendedKeyPair}. The former generates the RSA key pair for the signer according to the key generation parameters and the latter instantiates a new extended key pair with the default option of a geo-location graph encoding scheme.  The \texttt{ExtendedKeyPair} class also generates the bases for vertices and edges that will be used to encode the graph representation. For each key pair class there are two classes implementing the public and private keys. The \texttt{SignerPublicKey} class creates a new signer public key and also adds the contents of the public key to context of the challenge required in computing the challenge during proof execution. The private key of the signer is implemented with the \texttt{SignerPrivateKey} class. The extended public key instantiates a new extended public key attaching the bases, the graph encoding scheme and the graph encoding parameters. In addition it computes the current challenge context and setups the graph encoding.

There is a generic interface \texttt{IKeyPair} that returns the underlying base key pair. For instance, the \texttt{ExtendedKeyPair} will return the \texttt{SignerKeyPair}

\subsection{Utilities}

The utilities are placed in the \texttt{util} package. The main utilities in this package include classes implementing common functionality required from classes during the execution of proofs. The \texttt{Assert} implements common precondition checks for methods and constructors. Using this class we can evaluate if an object parameter is null or not, check if the string is empty and checking the length of a BigInteger number. There is also a utility class that displays and diagnoses the structure of a graph representation named \texttt{GraphUtils}. This class is used as a debugging tool during development. 

Another set of utility classes implement the URN scheme required for identifying elements in a proof and in the proof storage. There is also the \texttt{GSUtils} class that implements low-level cryptographic computations such as generating prime numbers and random numbers or computing the hash required during the computation of the challenge    

The \texttt{ProofStore} class implements a storage mechanism which is predominantly memory based. The goal of this class is to provide methods for storing and retrieving data related to proofs. Each element in the proof store is saved using its unique URN key identifier which can then be used to retrieve the information. 

\subsection{Graph Encoding}
The \GSL incorporates an interface for realizing graph encoding schemes called \texttt{IGraphEncoding}. It is an interface that is used to derive classes implementing encodings for graph representations.  The goal of the encoding is to map vertex and label strings to prime numbers that represent them in later stages of the graph signature scheme. An example encoding implementation is provided with the \GSL. The \texttt{GeoLocationGraphEncoding} class implements a graph encoding scheme that holds the geo-location of vertices using the structure of the UN ISO-3166~\cite{iso3166} alpha country codes. 



\begin{figure}[t]
	\begin{adjustbox}{center}
	\centering
		\includegraphics[width=7cm]{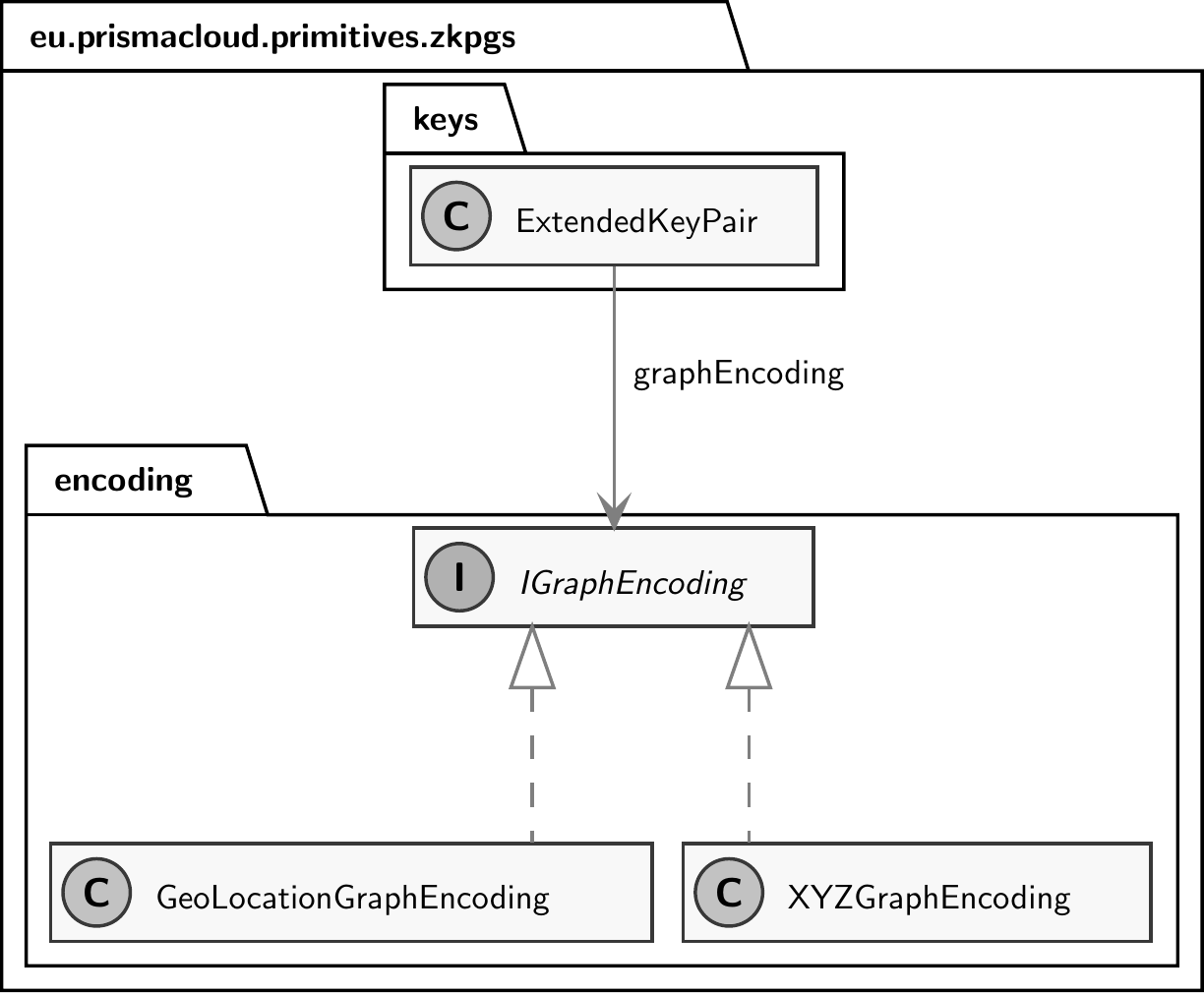}
	\end{adjustbox}
	\caption{Supporting arbitrary graph encoding schemes}
	\label{fig:uml-graph-encoding-scheme}
\end{figure}

\section{Evaluation}
\label{sec:evaluation}
In this section, we discuss the methods we used for evaluating the \GSL. First, we present the methods regarding testing and validation of the  graph signature library. Second, we discuss the results from the performance evaluation of the library.

\subsection{Library Testing and Validation }
The \GSL was tested and validated using a variety of methods. First, we tested the behaviour of each class and method following the unit testing~\cite{hunt2003pragmatic} paradigm. We also performed integration testing for each proof protocol that involves both parties. The validation of library consists of analyzing the source code using static analysis code tools and evaluating the dependencies of the library.

\paragraph{\textbf{Unit testing}}
We developed the test cases with the JUnit~\cite{junit} library version 5.5.2 to ensure the cryptographic algorithms and proof protocols are working correctly. There are 382 test cases that execute successfully. The test cases cover all the classes required to realize the graph signature scheme. The test coverage is $79\%$ of the source code.

Another feature that facilitates the testing of the \GSL is the introduction of a signing oracle that outputs graph signatures computed in a non-interactive way without the need for a recipient party. The main input to the signing oracle is a valid signer or extended key pair. The oracle can create graph signatures with a uniform random blinding randomness and a given graph representation encoded with a particular graph encoding scheme. According to the requirements of the test case the oracle can also create graph signatures on a group element or on a message $m_0$ without any graph encoding scheme or on a collection of bases. The flexible design of the signing oracle simplifies testing of the library without depending on the participation of the recipient to complete the graph signature.

\paragraph{\textbf{Static Code Analysis}}
We used a combination of three static code analysis tools to uncover critical bugs in the codebase. It has been shown that using multiple static code analysis tools can act complementary to each other for uncovering bugs in software~\cite{habib2018many}. SpotBugs\footnote{SpotBugs, https://spotbugs.github.io/}, Facebook's Infer~\cite{calcagno2015infer} and Google's error-prone~\cite{afta2012error} are the three tools for analyzing the \GSL codebase.

During the development and testing phase we analyzed the codebase using SpotBugs static code analysis tool version 3.1.12. This enabled us to improve the quality of the source code and avoid a variety of code smells.

The Infer static code analysis tool performs inter-procedural analysis and reports a variety of software issues such as null pointer exceptions and resource leaks. We have run this tool during the testing of the library to ensure there aren't any issues.

error-prone is build on top of javac compiler to check java code and presents the outcome of the static analysis as compilation errors. The advantage of this tool is that it reports only critical bugs without presenting a large number of warnings. In addition, error-prone results include a potential fix to the issue reported. We have successfully used this tool during the later stages of development and during testing to ensure that no critical bugs exists in the codebase of the library.

\paragraph{\textbf{Dependency Analysis}}
We used the OWASP dependency check~\cite{long2015depend} tool to evaluate the dependencies of the graph signature library and if they include any publicly disclosed known vulnerabilities. Currently, there aren't any vulnerable libraries bundled with the \GSL. This was achieved by including only a minimal set of dependencies in the library and implementing as much functionality as possible from scratch. Thus, we didn't require to add many libraries to realize the graph signature scheme.

\paragraph{\textbf{Information Flow}} We also tested that the classes used for the proof protocols do not leak any information during their execution that would compromise the security of the graph signature scheme. For instance, we want to ensure that the group elements do not leak the factorization of n. For this reason we first evaluate if the group element can be cast to a group element that would leak primes $p$ and $q$. Another check is if the group of the element leaks a group that knows the factorization of n. If the generator of the group leaks private information and if the group leaks the group order. These tests are contained in the \texttt{InfoFlowUtil} class.

We now discuss which specific parts of the library were tested for leakage of private information according to the above mentioned tests. First, the testing of the commitments component evaluates if the commitment value computed leaks any private information. Another check in this component is the testing of the base collection in a commitment that each group element that represents the base is assured that it does not leak any private information. If that is the case, then a malicious party could recreate the commitments and add his own information to the commitments so that she can compromise the proof protocols and in extension the graph signature scheme.

Second, the graph representation component includes the encoded bases which are used to encode the graph topology. We evaluate the encoded bases for information flow leakage by making sure that each group element that represents an encoded base does not leak private information. Third, the components for the extended and signer public key are tested for private information leakage. For the former component we test the auxiliary bases $S,Z,R$ and $R_0$, the certified bases for vertices and edges and the group for leaking private information. For the signer public key we evaluate the same items apart from the certified bases. Fourth, we ensure that the $A$ group element of the graph signature and the signed encoded bases do not leak private information. Fifth, for testing prover components we evaluate the group elements that represent witnesses in the proof protocols if they compromise the flow of information. Sixth, verification components are tested for information leakage by checking if bases, witnesses or the graph signature elements $A$ and $Q$ do not expose private information during the verification process.

\begin{knitrout}
\definecolor{shadecolor}{rgb}{0.969, 0.969, 0.969}\color{fgcolor}\begin{figure}

{\centering \includegraphics[width=0.8\textwidth]{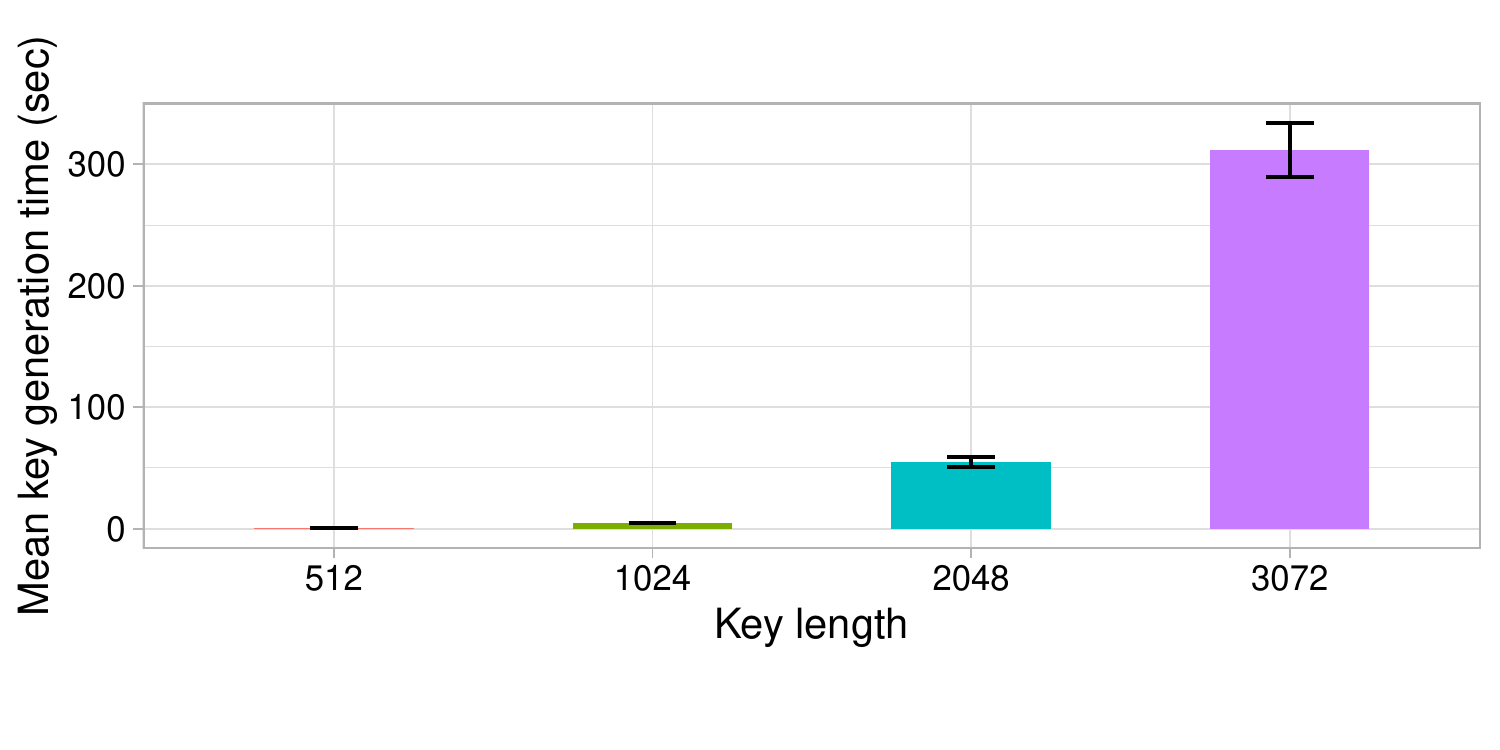} 

}

\caption[Mean time for generating the signer keys]{Mean time for generating the signer keys}\label{fig:keygen}
\end{figure}

\end{knitrout}Even though the information flow tests were executed during the testing stage of the \GSL, it provides assurances that the code developed does not leak any private information. However, it would be an improvement, if these checks could be executed during runtime using pre and post conditions. One way of achieving this is by using the Java Modeling Language (JML)~\cite{cok2011openjml}. JML is a behavioral interface specification language designed for Java. We consider developing JML specifications to prevent information flow leakages during runtime~\cite{scheben2016jmlinfo} as future work.

\subsubsection{Lessons learned}
The testing and validation of cryptographic libraries requires a lot of time and effort. In this subsection we discussed our approach to evaluate the \GSL. We believe that a variety of tools and methods should be used to not only test the correctness of the functionality but assess the fulfillment of security requirements as well. This enables to uncover as many bugs as possible using a variety of code analysis tools and ensure that private information will not leak. Another aspect of testing cryptographic libraries is to create components that can facilitate the testing of a cryptographic scheme. For instance, we have developed a graph signature oracle to rapidly test the issuing of graph signatures in a non-interactive way with various configurations.

\begin{figure}[tb]
  \begin{subfigure}[b]{0.8\textwidth}
    \includegraphics[width=\textwidth]{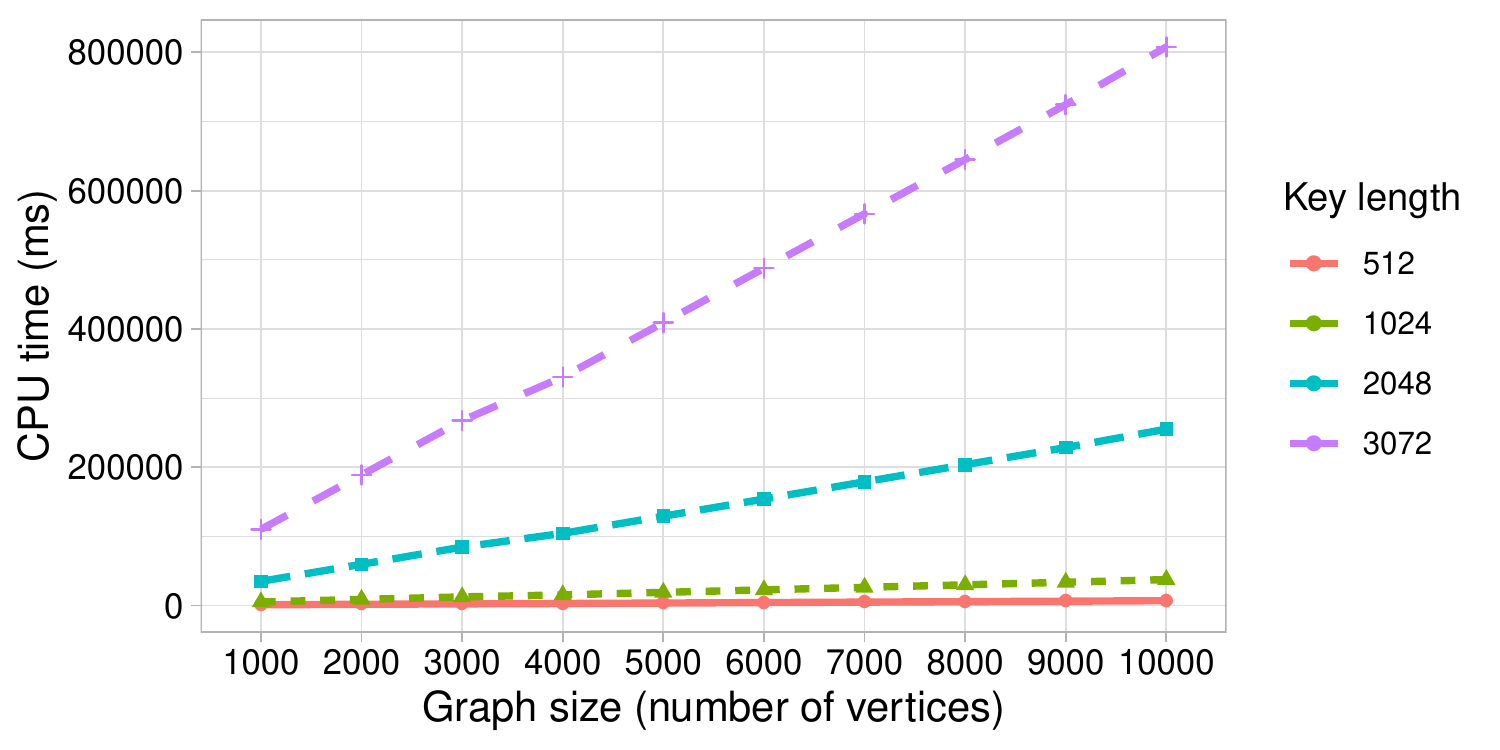}
	\caption{Mean CPU time for generating vertex and edge bases for the extended key generation}
	\label{fig:extendedkeygen1}
  \end{subfigure}
  \hfill
  \begin{subfigure}[b]{0.8\textwidth}
    \includegraphics[width=\textwidth]{figures/perf-extendedkeygen-1}
	\caption{Mean CPU time for the geolocation graph encoding setup}
	\label{fig:extendedkeygen2}
  \end{subfigure}
  \label{fig:extendedkeygen}
  \caption{Performance results for the extended key generation which involves generating new bases and setting up a graph encoding scheme}
\end{figure}\subsection{Performance Evaluation}
We evaluate the performance of the \GSL by adjusting the key length and the size of the graph.

\subsubsection{Experimental Setup}
We used a commodity desktop computer to execute the performance benchmarks. The testbed is equipped with a 3.4GHz Intel CoreTM i7–3770 CPU with 8 GB of RAM and a 500 GB hard disk. The operating system used to run the benchmarks is Ubuntu version 16.04. The Java Microbenchmarking Harness (JMH)~\cite{jmh} framework version 1.22 was used to develop the benchmarks. The YourKit\footnote{YourKit Java profiler, https://www.yourkit.com/java/profiler/} Java profiler version 2019.1 was used for measuring the CPU time for each benchmark.
Each test was executed at least 100 times unless other number is specified.

\subsubsection{Results}
\paragraph{\textbf{Key generation}}

\begin{figure}[tb]
  \begin{subfigure}[b]{0.8\textwidth}
    \includegraphics[width=\textwidth]{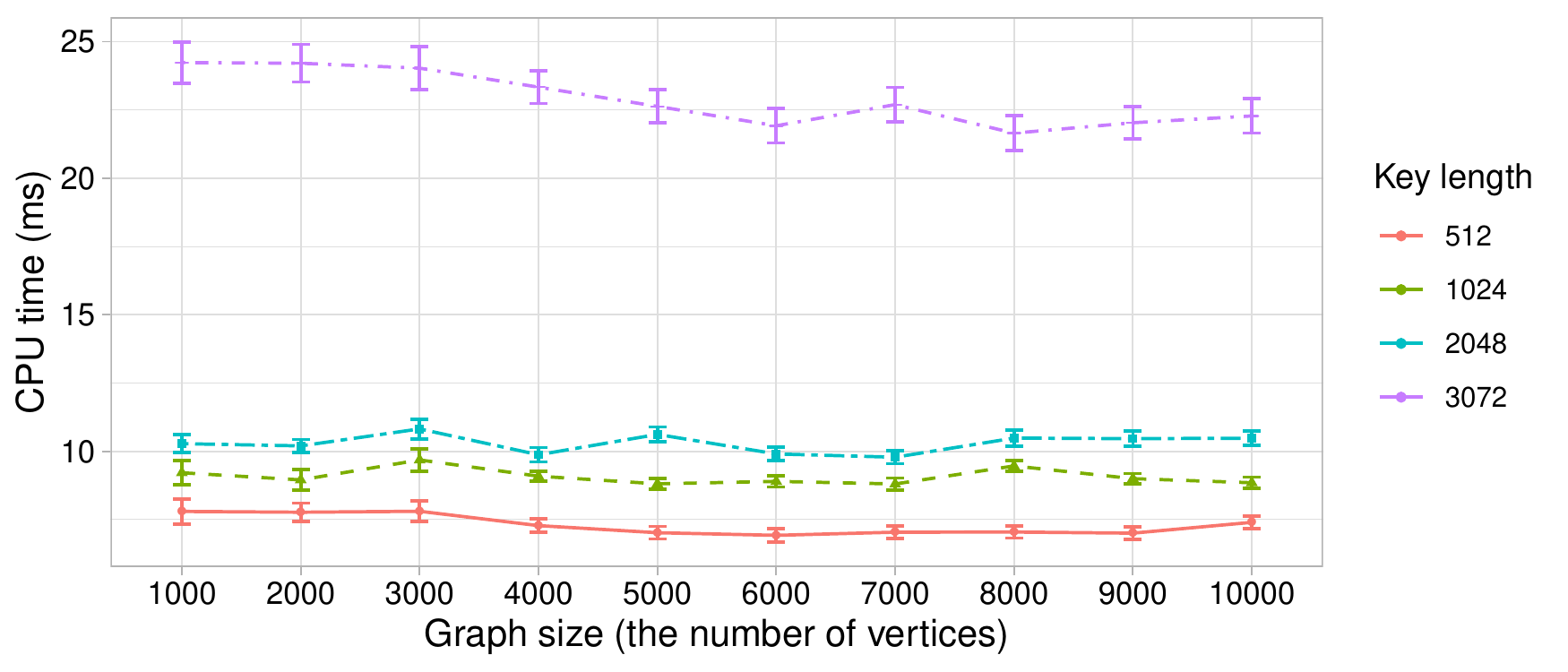}
	\caption{Mean CPU time for the signer to compute the partial signature}
	\label{fig:issuing_time-1}
  \end{subfigure}
  \hfill
  \begin{subfigure}[b]{0.8\textwidth}
    \includegraphics[width=\textwidth]{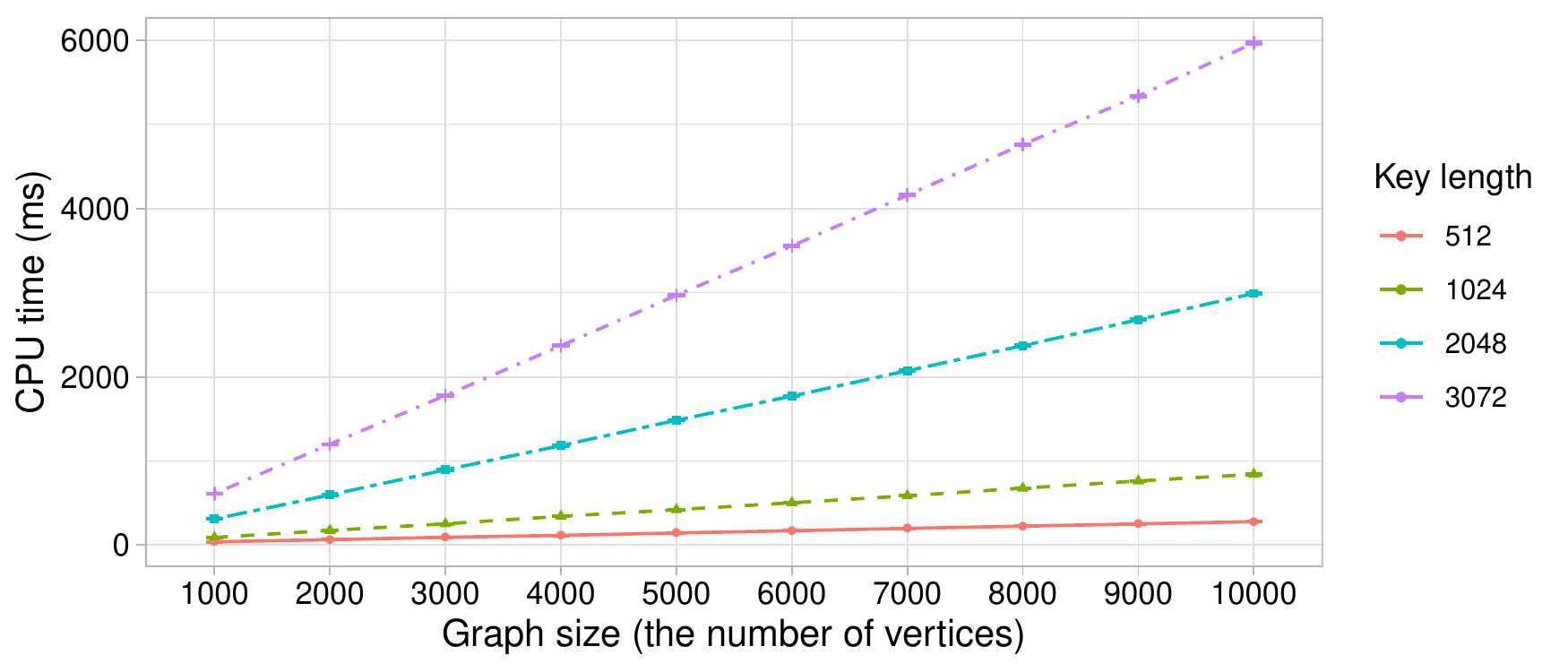}
	\caption{Mean CPU time for the recipient to complete the signature}
	\label{fig:issuing_time-2}
  \end{subfigure}
  \label{fig:issuing_time}
  \caption{Mean CPU time for issuing a graph signature for the signer and recipient}
\end{figure}

For the key generation experiment we measured how much time the graph signature library takes to generate a  key pair for the signer. The results are illustrated in Figure~\ref{fig:keygen} and are typical for creating RSA-based key pairs. The fastest option is using a 512-bit key followed by the 1024-bit key and the 2048 and 3072-bit key. The time that it takes to issue the graph signature is exponential to the bitlength of the key.

\paragraph{\textbf{Extended key generation}}

For this set of performance experiments we measured the time it takes to generate the required vertex and edge bases. These are the bases that are the placeholders for the encoding of the graph. In addition, we measure the time it takes to setup the geolocation graph encoding.
These two processes are part of the extended key generation.

%
%
%

\paragraph{\textbf{Issuing}} For this performance benchmark we measured the performance of the issuing of a graph signature. We first measure the time it takes the signer to issue the partial signature and the recipient to complete the signature. More precisely we measure the time it takes to compute the modular exponentiations for the issuing of the graph signature. The results are depicted in Figure 11.

%
%
%
%
%

The results show that when we sign a graph with a large number of vertices then the total issuing time is increased. We also see that the signer's issuing time is more than the recipient's for the same number of vertices and key length. This due to the fact that the recipient only completes the graph signature to create the final version of the signature. The key length directly affects the issuing time with a 512 key length having the best performance. The second best performance for the total issuing time is when the key length is 1024. The third best performance of the total issuing is when the key length is 2048. Using this key length is the minimum accepted configuration for using this library in a secure setting. We also see that this key length increases the issuing time in a higher degree than the issuing time for 512 and 1024 key lengths.

\paragraph{\textbf{Proof components}}

These tests adjust the length of the RSA key and measure the execution time of each proof component. We measure the mean CPU time for the main proof components in the graph signature scheme. First we measure the CPU time for the proof components used during the proving stage and then during the verifying stage. Figure~\ref{fig:proving} shows the time it takes to execute the commitment provers and the possession prover. Figure~\ref{fig:verifying} shows the time to execute the commitment verifier and the possession verifier.

\begin{knitrout}
\definecolor{shadecolor}{rgb}{0.969, 0.969, 0.969}\color{fgcolor}\begin{figure}

{\centering \includegraphics[width=0.99\textwidth]{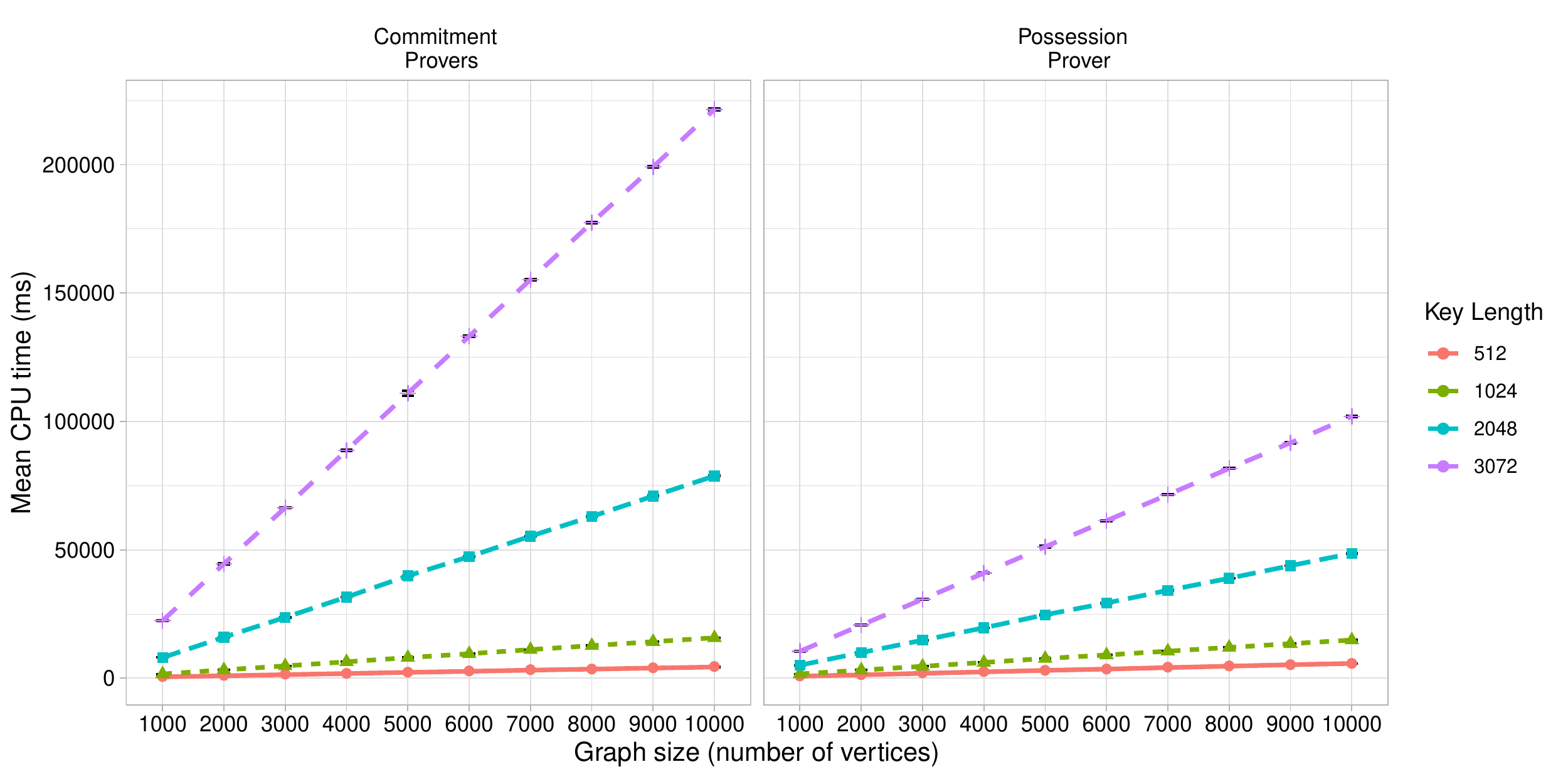} 

}

\caption[Mean CPU time for the  Commitment and Possession provers using different key lengths and graph size]{Mean CPU time for the  Commitment and Possession provers using different key lengths and graph size}\label{fig:proving}
\end{figure}

\end{knitrout}

\begin{knitrout}
\definecolor{shadecolor}{rgb}{0.969, 0.969, 0.969}\color{fgcolor}\begin{figure}

{\centering \includegraphics[width=0.99\textwidth]{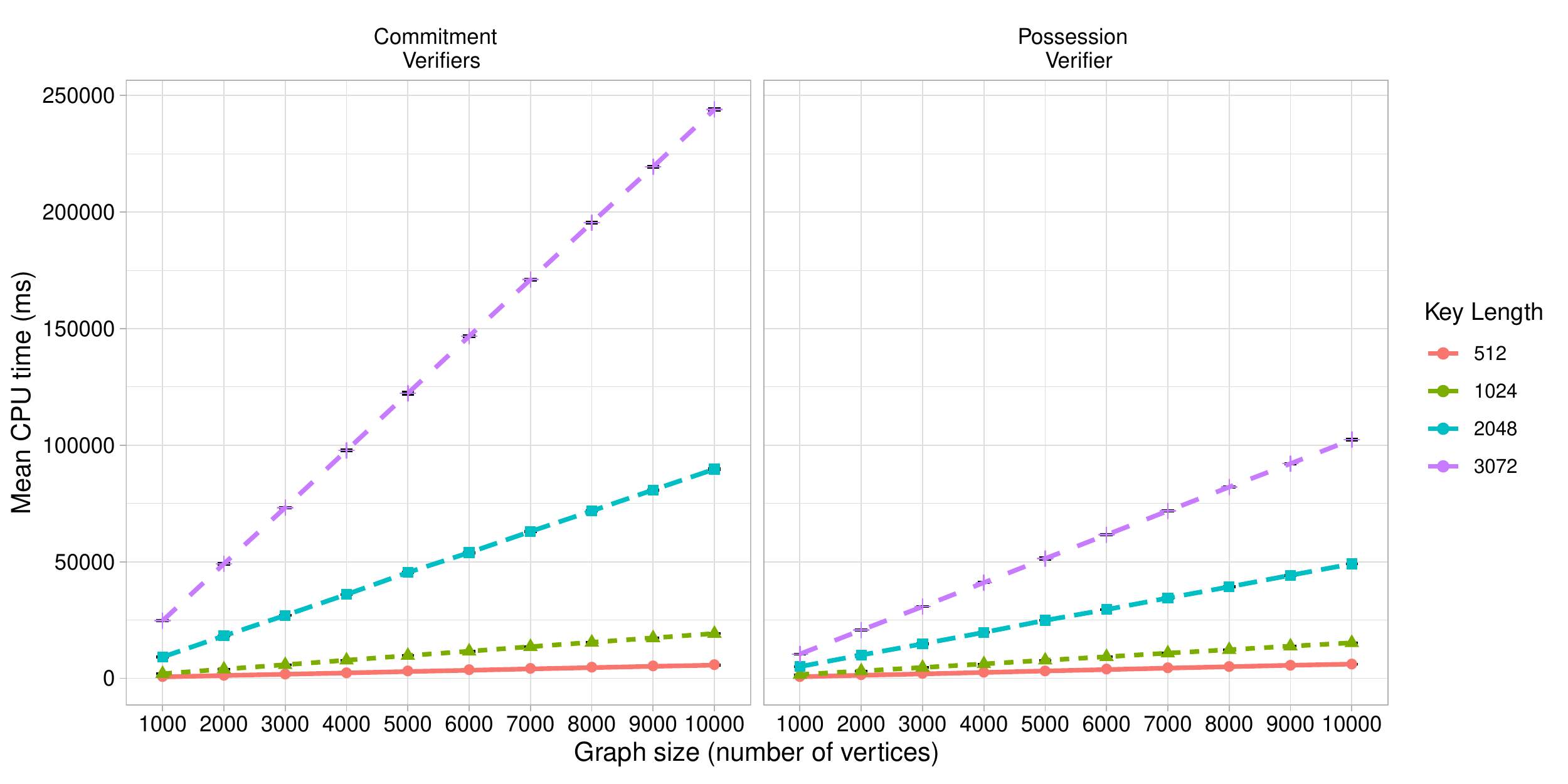} 

}

\caption[Mean CPU time for the Commitment and Possession verifiers using different key lengths and graph size]{Mean CPU time for the Commitment and Possession verifiers using different key lengths and graph size}\label{fig:verifying}
\end{figure}

\end{knitrout}


\section{Related Work}
\label{sec:relatedwork}


\paragraph{\textbf{Cryptographic libraries}} In recent years cryptographic libraries have been introduced that either support a wide variety of cryptographic primitives or support only a limited set of cryptographic primitives. In the first category of cryptographic libraries includes the OpenSSL library. OpenSSL~\cite{openssl} provides a complete implementation of cryptographic primitives. cryptlib~\cite{gutmann:cryptlib} is an early attempt of a modular cryptographic library which offers a generic interface and implements a wide variety of cryptographic primitives. Other libraries that implement multiple cryptographic schemes include NaCl~\cite{bernstein2012security} and Apache Milagro~\cite{milagro:crypto}.

The second category of cryptographic libraries include libraries that support a specific cryptographic primitive. For this work, we are interested in cryptographic libraries that support anonymous credentials and especially attribute-based anonymous credentials that are related to the graph signature scheme. Identity mixer (IDEMIX)~\cite{camenisch2002design} implements anonymous credentials and zero-knowledge proofs of knowledge. The Charm framework~\cite{Akinyele2013} provides a method to rapidly prototype cryptographic systems. Other cryptographic libraries focus on implementing anonymous credentials. The CLARC library implements anonymous credentials with a reputation system ~\cite{Bemmann2018clarc}. The IRMA platform~\cite{alpar2017irma,brands2000uprove} implements the IDEMIX attribute-based credential scheme for android mobiles. The main shortcoming of this work is that the implementation is not easily portable to other platforms, since it was not designed as a reusable library. The U-Prove~\cite{paquin:uprove:2013,brands2000uprove} system by Microsoft implements attribute-based anonymous credentials. Emmy~\cite{stopar2019emmy} is an open-source cryptographic library that focuses on trust-enhancing authentication for online services.
ZKPDL~\cite{mekhl2010zkpdl} is a language-based system for zero-knowledge proofs that implements anonymous credentials for e-cash.

\section{Concluding remarks and future work}
\label{sec:conclusion}
In this work we have introduced the design and implementation of a cryptographic scheme in a cryptographic library. We discussed the design and architecture of the library that realizes the graph signature scheme and discussed the reasons behind our design decisions. During the course of the development of this library we learned valuable lessons that can be beneficial for other developments of cryptographic primitives. We believe that discussing the lessons learned will assist developers in improving their design and architecture of cryptographic libraries. The evaluation results show that the library performs on par with other RSA-based signature schemes. Each proof component exhibit performance that was expected.

Future work entails the integration of an elliptic curve subsystem that would enable the library to realize a graph signature scheme based on elliptic curves. Such an integration will truly test the design and architecture of the library. We intend to evaluate the performance of the elliptic curve subsystem with the RSA one.

\section*{Acknowledgement}
This work was supported in full by the \CASCAde. 
\bibliography{bibliography}
\bibliographystyle{plain}


\end{document}